\documentclass[conference]{IEEEtran}
\IEEEoverridecommandlockouts
\usepackage{cite}
\usepackage{amsmath,amssymb,amsfonts}
\usepackage{algorithmic}
\usepackage{graphicx}
\usepackage{textcomp}
\usepackage{xcolor}

\usepackage{epsfig,graphicx}
\usepackage{nomencl}
\makenomenclature
\usepackage{array}
\usepackage{lscape}
\usepackage{float}
\usepackage{color}
\usepackage{algorithm}
\usepackage{algorithmic}
\usepackage{amssymb}
\usepackage{subfigure}
\usepackage{multirow}

\usepackage{amsmath,graphicx,amsthm}
\usepackage{amssymb}
\usepackage{epstopdf}
\usepackage{subfigure}
\usepackage{amsfonts}
\usepackage{hyperref}
\hypersetup{nolinks=true}
\usepackage{subfigure}

\usepackage{color,soul}
\usepackage{multicol}

\usepackage{lipsum}
\usepackage{mathtools}
\usepackage{cuted}

\def\BibTeX{{\rm B\kern-.05em{\sc i\kern-.025em b}\kern-.08em
    T\kern-.1667em\lower.7ex\hbox{E}\kern-.125emX}}
\begin{document}

\title{On the Impact of CDL and TDL Augmentation for RF Fingerprinting under Impaired Channels\\
}

\author{ \IEEEauthorblockN{Omer Melih Gul$^{\dag}$, Michel Kulhandjian$^{\dag}$,  Burak Kantarci$^{\dag}$, Claude D'Amours$^{\dag}$, Azzedine Touazi$^{\ddag}$, Cliff Ellement$^{\ddag}$
		}
		
			\IEEEauthorblockA{$^{\dag}$Department of EECS,
			University of Ottawa,
			Ottawa, ON, Canada \\
			E-mail: \texttt{\{ogul,mkulhand,burak.kantarci,cdamours\}@uottawa.ca}  \\
			$^{\ddag}$Artificial Intelligence Solutions, ThinkRF, Ottawa, ON, Canada\\
			E-mail: \texttt{\{azzedine.touazi,cliff.ellement\}@thinkrf.com }  } }
\maketitle

\begin{abstract}
Cyber-physical systems have recently been used in several areas (such as connected and autonomous vehicles) due to their high maneuverability. On the other hand, they are susceptible to cyber-attacks. Radio frequency (RF) fingerprinting emerges as a promising approach. 
This work aims to analyze the impact of decoupling tapped delay line and clustered delay line (TDL+CDL) augmentation-driven deep learning (DL) on transmitter-specific fingerprints to discriminate malicious users from legitimate ones.
This work also considers 5G-only-CDL, WiFi-only-TDL augmentation approaches.
RF fingerprinting models are sensitive to changing channels and environmental conditions. For this reason, they should be considered during the deployment of a DL model. Data acquisition can be another option. Nonetheless, gathering samples under various conditions for a train set formation may be quite hard. Consequently, data acquisition may not be feasible.
This work uses a dataset that includes 5G, 4G, and WiFi samples, and it empowers a CDL+TDL-based augmentation technique in order to boost the learning performance of the DL model.
Numerical results show that CDL+TDL, 5G-only-CDL, and WiFi-only-TDL augmentation approaches achieve 87.59\%, 81.63\%, 79.21\% accuracy on unobserved data while TDL/CDL augmentation technique and no augmentation approach result in 77.81\% and 74.84\% accuracy on unobserved data, respectively. 
\end{abstract}
    
\begin{IEEEkeywords}
Data augmentation, deep learning, RF fingerprinting, secure design, uncrewed aerial vehicles
\end{IEEEkeywords}


\section{Introduction}
Cyber-physical systems which are networked systems with physical and computational capabilities can be applied to many areas. They include wireless communications, search and rescue, crop spraying and monitoring, and agricultural, fire, or wildlife surveillance. 
Their integration with wireless communication capabilities brings many advantages. Nevertheless, they may face security issues. To name a few for a sample cyber-physical system, an uncrewed aerial vehicle (UAV) may be susceptible to tampering, masquerade, spoofing, and replay attacks \cite{Yaacoub,Alladi}.

By considering their various features, security solutions are suggested at various levels for these systems. An intrusion detection system is used to identify suspicious activities in application and network layers. Furthermore, applying message encryption at different levels is of paramount importance to protect data at rest and in transmission. On the basis of this motivation, authentication and encryption-based security solutions for cyber-physical systems have been widely investigated \cite{Yaacoub}. Recently, Alladi \textit{et al.} \cite{Alladi} introduce an authentication system based on a physical unclonable function. Bassey \textit{et al.} \cite{Bassey} discuss the use of authentication codes for an Internet of Things (IoT) device, and by leveraging the Kolmogorov-Smirnov test, it determines the legitimacy of the user of an IoT device.

Despite the existence of encryption and authentication techniques, cyber-physical systems are vulnerable to spoofing attacks that can be coped with if system-level security complements encryption and authentication-based solutions. 
For instance, Shi \textit{et al.} \cite{Shi} present an anticipatory study via a generative adversarial network (GAN)-based spoofing attacks that aim to bypass the authentication barrier by mimicking legitimate transmitters. To do so, the GAN-based adversarial attack captures channel effects and waveforms. Consequently, it increases the success probability of such an attack. GANs can also be used to safeguard systems against spoofing attacks. Sagduyu \textit{et al.} \cite{Sagduyu} exploit the classification outputs of a neural network to deceive adversaries. At lower layers such as in the radio frequency (RF) domain, security by design can be offered to cyber-physical systems.

RF security entails hardware-specific characteristics and imperfections of transmitters. For example, in a connected vehicle network, each vehicle transmitter reveals hardware imperfections during communications via RF waves (at physical layer), which is referred to as RF fingerprint of their transmitters. For ultra-reliable low latency communications (URLLC) purposes, RF fingerprinting can suggest security solutions to accelerate security services in comparison to network-layer (L3) solutions. RF fingerprinting can be defined as mining hardware imperfection patterns in manufacturing defaults \cite{Shiu}. DL methods are widely studied and shown to be successful at extracting specific features. Therefore, DL models can be utilized to detect particular RF signatures \cite{Karunaratne}.   

Channel impairments affect the performance of DL techniques to identify the RF signatures of a transmitter \cite{Jagannath}, \cite{RFChapter}. 
It is not feasible to collect data in all circumstances. For this reason, applying augmentation methods for RF fingerprinting is a viable solution to deal with channel impairments during the DL training process. As a special type of augmentation technique, adversarial machine learning can be used prior to training DL classifiers for RF fingerprinting \cite{Kantarci2022}.


This study that was initially presented in \cite{48WWRF} differs from the existing works with the following points:
\begin{itemize}
\item To the best of our knowledge, this is the first work focusing on applying different types of data augmentation on raw time-domain signals in heterogeneous environments with real-world data. 
\item We decouple the tapped delay line (TDL) and clustered delay line (CDL) augmentation approaches such that 5G data is augmented by CDL whereas WiFi data is augmented by transforming it by TDL (referred to as CDL+TDL hereafter). 
\end{itemize}

This study extends our work in \cite{48WWRF} with the following points:
\begin{itemize}
\item We propose a 5G-only-CDL augmentation approach where only 5G data is augmented with CDL transform. Thus, we can observe its difference from CDL/TDL augmentation.
\item We propose a WiFi-only-TDL augmentation approach where only WiFi data is augmented with TDL transform. Thus, we can observe its difference from CDL/TDL augmentation.
\end{itemize}

This paper is structured as follows. Related work is reviewed in Section \ref{sec:rel}. Section \ref{sec:sys_pro} gives the system model and problem definition. 
Section \ref{sec:app0} introduces the 5G-only-CDL and WiFi-only-TDL augmentation techniques.
Section \ref{sec:app} introduces the decoupled CDL+TDL augmentation technique. Section \ref{sec:num} presents simulation results. Section \ref{sec:con} concludes this work and presents future work.

\section{Related Work} \label{sec:rel}
This section presents the most relevant literature on RF fingerprinting. The work presented by Cekic \textit{et al.} primarily focuses on multiple WiFi protocols (IEEE 802.11a and IEEE802.11.g) and automatic dependent surveillance-broadcast (ADS-B) protocol \cite{Cekic} to identify two main transmitter characteristics, namely in-phase and quadrature-phase (I/Q) imbalance, and power amplifier nonlinearity. A comparison of additive white Gaussian noise (AWGN) and without augmentation is presented. Shea \textit{et al.} in \cite{Shea} report that complex-value networks may not be sufficient for radio signal classification in many real-world applications.

\begin{table}
\centering
\caption{Brief comparison of techniques proposed in close literature}
\label{tab:ConfigurationTypes}
\begin{tabular}{|p{1.2cm}|p{1cm}|p{1cm}|p{0.9cm}|p{1.5cm}|}
\hline
 & Demodu- lated or real data   & Different day tests & Multiple waveforms & CDL\&TDL augmentation analysis\\
\hline
Cekic \textit{et al.} \cite{Cekic}     & no  & yes & yes & no \\
\hline
Sankhe \textit{et al.} \cite{Sankhe}   & no  & yes & no & no \\
\hline
Mohanti \textit{et al.}\cite{Mohanti}  & no  & yes & no & no \\
\hline
Soltani \textit{et al.} \cite{Soltani} & no  & yes & no & no \\
\hline
Muns \textit{et al.} \cite{Muns}       & yes  & yes & yes & no \\
\hline
Zhao \textit{et al.} \cite{Zhao}       & yes  & no & no & no \\
\hline
Comert \textit{et al.} \cite{Comert}   & yes  & yes & yes & no \\
\hline
Gul \textit{et al.} \cite{48WWRF}      & yes  & yes & yes & yes\\
\hline
This work                    & yes  & yes & yes & yes\\
\hline
\end{tabular}
\label{table1}
\end{table}

Sankhe \textit{et al.} \cite{Sankhe} leverage raw I/Q samples to characterize static channels, which eliminates the coefficient prediction requirement. On the other hand, complex demodulated symbols are utilized to remove channel effects. By exploiting direct current (DC) offset and I/Q imbalance, a convolutional neural network (CNN) is trained to recognize radio signal signatures under static and varying channel characteristics. 

Ozturk \textit{et al.} \cite{Ozturk} present a CNN-based approach that takes spectrograms and time-series images of RF signals as inputs. Transient signals with low signal-to-noise-ratio (SNR) are prone to noise effects; hence, it is not viable to use a time series-based model. As a result of the frequency range consideration by spectrograms, more accurate outputs can be obtained in comparison to the time series-based model.

RF signatures of UAVs are aimed to be identified by Mohanti \textit{et al.} \cite{Mohanti} under a CNN with VGG backbone. To tackle I/Q imbalance, amplitude changes are fed into real and imaginary components. 
It is worth noting that DL-based fingerprinting mechanisms can recognize UAV radio signals only if the channel conditions do not change across training and test phases. To cope with this phenomenon, the authors propose a processing block to arrange I/Q instances.

Soltanie \textit{et al.} \cite{Soltani} propose two alternate augmentation methods that build on two finite impulse response filters. The first method is trained under a dataset generated in MATLAB and validated under a DARPA dataset. These methods have been shown to run with raw I/Q instances without quadrature phase shift keying (QPSK) modulation. 
The high throughput task group (TGn) of wireless local area network (WLAN) channel data samples are augmented with AWGN to mimic more realistic channel behavior. This method can perform well with channels that have not been seen before. Testing of this method is performed by considering the boundaries of single-channel models. 

For RF fingerprinting purposes, Reus-Muns \textit{et al.} \cite{Muns} built a dataset that comprises WiFi, 5G, and LTE samples and feed raw I/Q samples into a neural network. The focal point of that work is to investigate the accuracy degradation between training and test data acquired on different days with different channel conditions. When the train and test data belong to the same day, the accuracy is sufficient whereas when the test is performed in a different day, the accuracy is remarkably degraded. To remediate this issue, leveraging triplet loss functions is proposed by Reus-Muns \textit{et al.} \cite{Muns} so that the accuracy is acceptable even if the training and test data are collected on different days, i.e., with different propagation conditions. Zhao \textit{et al.}  \cite{Zhao} present an analysis of wireless signals for UAV model classification by combining auxiliary classifier generative adversarial network (ACGAN) and Wasserstein generative adversarial network (WGAN) where root mean squared propagation is applied as an accelerator and signals obtained by applying a band-pass filter.

Table \ref{table1} presents a comparative list of the relevant works that are closest to the work presented in this paper. It is observed that the majority of the relevant works rely on either synthetic datasets generated in MATLAB or datasets that contain modulated data. Training and testing on different days (i.e., varying channel conditions) is pursued by few studies. Since CDL/TDL augmentation was shown to perform well in the previous work in \cite{Comert}, this work aims to pursue a detailed investigation of decoupling CDL and TDL on 5G and WiFi waveforms, respectively under data-scarce scenarios.

\section{System Model and Problem Definition}  \label{sec:sys_pro}
\begin{figure}[h]
    \includegraphics[width=0.98\linewidth]{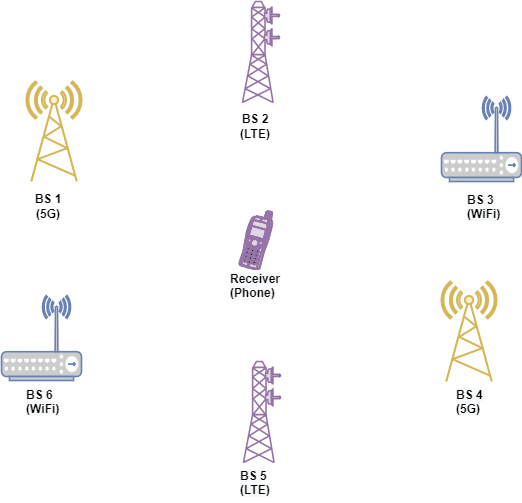}
    \caption{In this network, base stations B5 and B6 communicates with mobile phone by resembling B2 and B3 (spoofing attack).}
    \label{System_Model}
\end{figure}

This section introduces the system model before proceeding with the methodological details. Figure \ref{System_Model} illustrates the system model.

Autonomous vehicles, similar to other cyberphysical systems, are vulnerable to spoofing attacks.
Each wireless device operates with slight hardware imperfections that are caused during manufacturing. These hardware imperfections can be recognized on radio signals as their wireless signatures, which are referred to as the RF fingerprint of the device. Alongside the enhanced mobile broadband (eMBB) and massive machine type communications (mMTC), the ultra-reliable low latency communications (URLLC) use case of 5G, requires security by design, for which RF fingerprinting is an ideal candidate.

The problem formulation for RF fingerprinting is as follows: A receiver aims to identify transmitters based on their RF signatures in time-domain raw signal samples.

The objective of this study is to obtain high accuracy under Day-2 data when the radio signal classifier is trained under Day-1 data. Thus, we aim to provide insights into decoupling the channel condition effects and the transmitter impairments. The radio signal instances in the data consist of 5G, WiFi, and LTE waveforms in the POWDER Dataset that were originally presented in \cite{Muns}. The dataset includes 60 .bin files for Day-1 and 60 .bin files for Day-2 that are equally split between the 5G, 4G/LTE, and WiFi waveforms for both days.

\section{5G-only-CDL and WiFi-only-TDL Augmentation Approaches}  \label{sec:app0}

As previously said, RF fingerprinting is susceptible to a variety of channel and environmental conditions; as a result, RF fingerprinting asks for effective solutions that are robust (in terms of accuracy) in a variety of circumstances. Data augmentation-based solutions are required to develop neural network architectures that take the channel conditions into consideration. In fact, the goal of data augmentation is to develop a method that can replicate radio signal traces in a variety of channel and environmental settings.

In this section, we propose two augmentation approaches where each of CDL and TDL transforms are applied to just one of the communication technologies data (5G, WiFi) instead of applying the same transform to all data.
In the first subsection, we first propose a 5G-only-CDL augmentation approach different from CDL/TDL approach.
Then, in the second subsection, we propose a WiFi-only-TDL augmentation approach different from CDL/TDL approach.

\subsection{5G-only-CDL Augmentation Approach}  \label{sec:app01}

In this subsection, we would like to observe the effect of applying CDL transform on just a subset (5G data) of the dataset instead of applying CDL transform to all data. This provides more insight for us to present the CDL+TDL augmentation approach, which is proposed in \cite{48WWRF}, in the following section.

\begin{figure}[t]
    \includegraphics[width=0.73\linewidth]{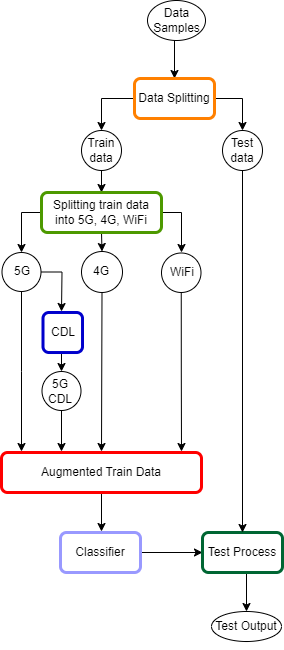}
    \caption{The proposed 5G-only-CDL augmentation architecture}
    \label{HighLevelModel_5GCDL}
\end{figure}

We suggest using the MATLAB 5G Toolbox (found in \cite{Mathworks}) to apply CDL-transform on only 5G signals for data augmentation to train the model before the unseen data with channel impairments. A minimalist illustration of 5G-only-CDL augmentation is given in Figure \ref{HighLevelModel_5GCDL}. 

\subsection{WiFi-only-TDL Augmentation Approach}  \label{sec:app02}

In this subsection, we would like to observe the effect of applying TDL transform on just a subset (WiFi data) of the dataset instead of applying TDL transform to all data. This provides more insight for us to present the CDL+TDL augmentation approach, which is proposed in \cite{48WWRF}, in the following section.

We suggest using the MATLAB 5G Toolbox (found in \cite{Mathworks}) to apply TDL-transform on only WiFi signals for data augmentation to train the model before the unseen data with channel impairments.

\begin{figure}[t]
    \includegraphics[width=0.73\linewidth]{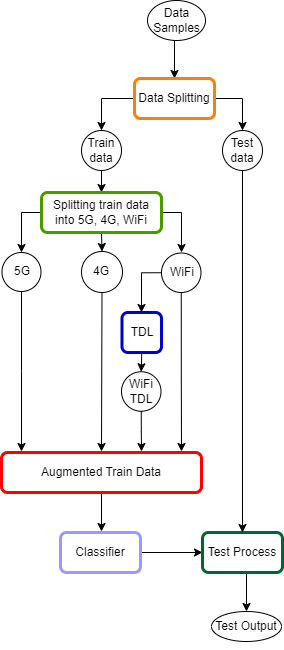}
    \caption{The proposed WiFi-only-TDL augmentation architecture}
    \label{HighLevelModel_WiFiTDL}
\end{figure}

A minimalist illustration of WiFi-only-TDL augmentation is given in Figure \ref{HighLevelModel_WiFiTDL}.

\section{Decoupled CDL and TDL (CDL+TDL) Augmentation Approach}  \label{sec:app}

As stated earlier, RF fingerprinting is vulnerable to varying channel and environmental conditions; hence, RF fingerprinting calls for effective solutions that are robust (in terms of accuracy) under varying conditions. To account for the channel conditions in the design of a neural network architecture, data augmentation-based solutions are needed. Indeed, the rationale for data augmentation is to come up with a technique that can mimic radio signal traces under varying channel and environmental conditions. 

In order to train the model prior to the unobserved data with channel impairments, we propose to decouple CDL and TDL augmentation via MATLAB 5G Toolbox  (available in \cite{Mathworks}) for 5G and WiFi signals. All models are designed for full frequency ranges. It may be scaled further for observing the desired root-mean-square (RMS) delay spread. 
Therefore, special models for line-of-sight (LOS) and non-line-of-sight (NLOS) scenarios can be designed \cite{etsi}. 
In addition to the Day-1 data of the dataset, we aim to add more knowledge that represents characteristics of unobserved data on Day-1. The rationale for this is to cope with the  channel impairments / varying channel conditions on Day-2. To do so, more information is added to Day-1 data so as to learn and handle channel impairments during the test of Day-2 data. 

The previous work in \cite{Comert} introduces CDL/TDL augmentation which is quite suitable to output augmented data from 5G samples. Additionally,  this paper shows that augmenting WiFi samples with TDL works well with WiFi samples.

This work introduces the decoupled CDL+TDL augmentation technique that augments 5G data samples by transforming them via CDL transformation, and WiFi data samples by transforming them via TDL transformation. Day-1 (train) data is augmented via this augmentation technique, and Day-2 data for testing.

It should be noted that CDL/TDL augmentation technique applies the same transform to all waveforms types of 5G, LTE, and WiFi data. In other words, it applies CDL transform to all or TDL transform to all. Both transforms are highly correlated with each other so their performance figures are alike. On the other hand, the CDL+TDL augmentation technique decouples TDL and CDL transforms by applying CDL transform just on 5G-waveform data and TDL transform just on WiFi data for data augmentation (LTE-waveform data is not augmented in CDL+TDL augmentation approach). 

\begin{figure}[t]
    \includegraphics[width=0.70\linewidth]{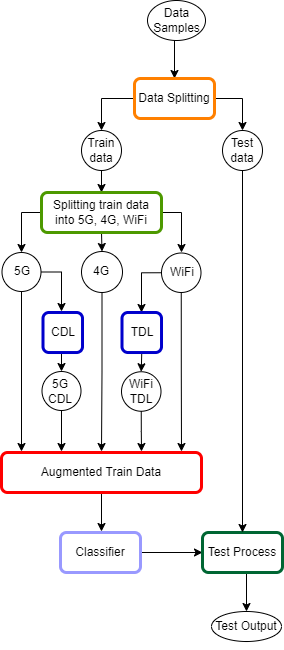}
    \caption{The proposed decoupled CDL+TDL augmentation architecture}
    \label{HighLevelModel_CDL+TDL}
\end{figure}

A minimalist illustration of decoupled CDL+TDL augmentation is given in Figure \ref{HighLevelModel_CDL+TDL}.

\begin{figure}
    \includegraphics[width=0.7\linewidth]{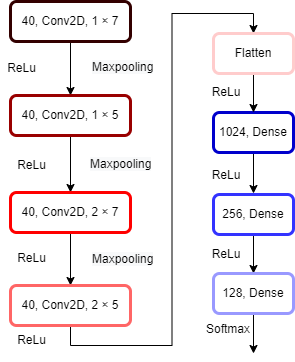}
    \caption{The classifier used in CDL+TDL augmentation architecture 
    }
    \label{classifer}
\end{figure}

Figure \ref{classifer} shows the classifier structure which is adopted from the study in \cite{Muns}.

\section{Numerical Results} \label{sec:num}
As mentioned in Section III, the POWDER dataset is used in this work \cite{Muns} to test the decoupled CDL+TDL, 5G-only-CDL, and WiFi-only-TDL augmentation approaches. The public dataset used in this work includes raw I/Q signal samples transmitted from four base stations. The data instances of 5G, WiFi, and LTE signals in the dataset are collected on two different days, i.e., Day-1 and Day-2.

To cope with the channel impairments, we use CDL transformation for 5G samples and TDL transformation for WiFi samples for data augmentation.  Moreover, for a better comparison of CDL+TDL, 5G-only-CDL, and WiFi-only-TDL augmentation approaches with CDL/TDL augmentation introduced in \cite{Comert}, we use the same settings in \cite{Comert} and the POWDER dataset in \cite{Muns}.

Table \ref{table2} exhibits the accuracy levels under the CDL+TDL augmentation technique, 5G-only-CDL, WiFi-only-TDL, TDL/CDL augmentation technique, and no augmentation in the four base station scenario under the POWDER dataset. The training process lasts $16$ epochs for all cases. 

\begin{table}[h]
\centering
\caption{Comparison of accuracy levels obtained on Day-1 and Day-2}
\label{tab:ConfigurationTypes2}
\begin{tabular}{|p{2cm}|p{2.75cm}|p{2.75cm}|}
\hline
Augmentation Method & Accuracy on Day-1 (\%)  & Accuracy on Day-2  (\%)  \\
\hline
Without Augmentation  & 99.72 & 74.84   \\
\hline
TDL/CDL-based Augmentation         & 99.30 & 77.81   \\
\hline
5G-only-CDL Augmentation         & 99.20 & 81.63   \\
\hline
WiFi-only-TDL Augmentation         & 99.23 & 79.21   \\
\hline
CDL+TDL Augmentation           & 97.35 & 87.59   \\
\hline
\end{tabular}
\label{table2}
\end{table}

\begin{figure}
\centering
    \includegraphics[width=\linewidth]{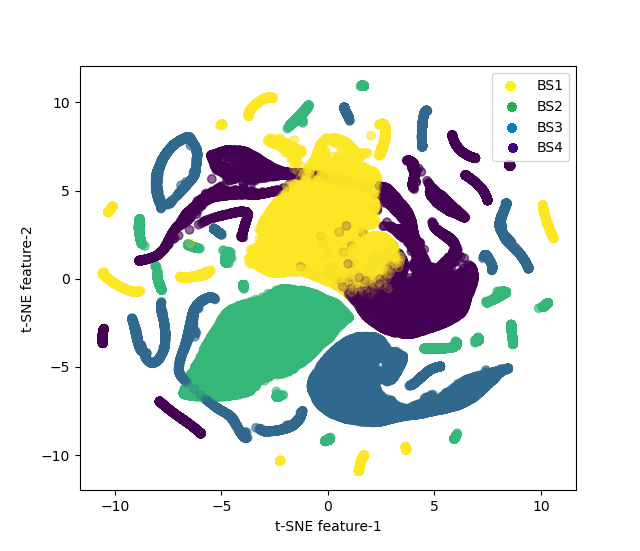}
    \caption{t-SNE visualization for Train Data with no augmentation}
    \label{fig:zero}
\end{figure}

\begin{figure}
\centering
    \includegraphics[width=\linewidth]{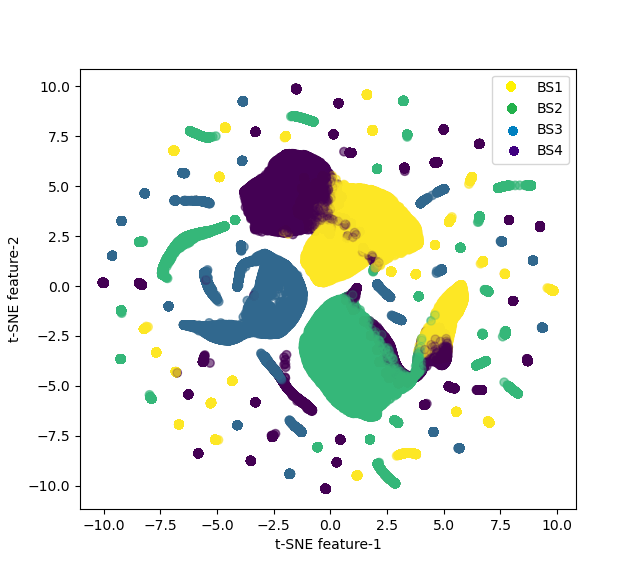}
    \caption{t-SNE visualization for Train Data with TDL/CDL Augmentation}
    \label{fig:first}
\end{figure}

\begin{figure}
    \includegraphics[width=\linewidth]{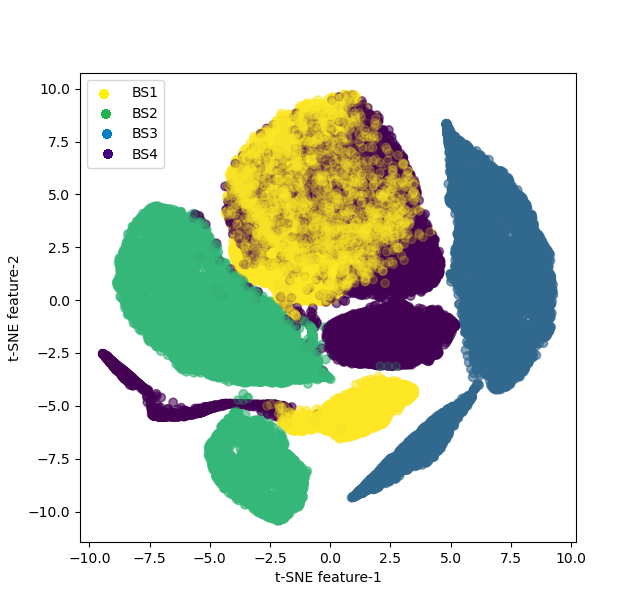}
    \caption{t-SNE visualization for Test data
    }
    \label{fig:second}
%
\label{fig:TSNE}
\end{figure}

Numerical results show that the decoupled CDL+TDL augmentation achieves significantly better performance than the TDL/CDL-based augmentation and no augmentation approaches for the scenario where the model is trained with Day-1 data and tested under Day-2 data. According to the numerical results, the decoupled CDL+TDL, 5G-only-CDL, WiFi-only-TDL augmentation techniques achieve an accuracy level of 87.59\%, 81.63\%, 79.21\%, respectively on previously unobserved data whereas TDL/CDL augmentation technique and no augmentation approach result in an accuracy level of 77.81\% and 74.84\%, respectively on unobserved data.
Figures \ref{fig:zero}-\ref{fig:second} demonstrate the t-SNE plots for four base stations to visualize the impact of the augmentation methods.


\section{Future Research Challenges} \label{sec:fut}

The paper investigates wireless security problems and proposes wireless device fingerprinting methods. In this section, we identify a few significant open research difficulties and possibilities to construct a robust radio frequency fingerprinting system (RFFS) for completeness. These issues include simulation-reality gap, the impact of receiver hardware, RFFS vulnerabilities \cite{Jagannath}.

\subsection{Simulation-reality gap}

Synthetic data realism is also crucial. Deep learning models trained on synthetic data are hard to apply to real radio transmissions. The synthetic dataset's assumptions about transmitter hardware defects and fading channel vs actual hardware and environmental impacts create a capability gap. Real-world data from IoT sensors and radios are scarce, which leads to synthetic data. In natural language processing (NLP) and computer vision (CV), large-scale datasets like MNIST \cite{151}, Stanford sentiment \cite{152}, IMDb \cite{153}, Sentiment140 \cite{154}, etc., are easily available. Existing datasets cannot be used in different machine learning frameworks since there is no standard dataset structure and organization. To achieve universal performance, neural networks must be trained with more data. Generalization is the first step to deployment-ready fingerprinting.

\subsection{Impact of receiver hardware} 

Like transmitter hardware, receiver hardware captures and processes emissions for fingerprinting. The receiver hardware's phase noise, clock offsets, filter distortions, IQ imbalance, etc., could alter the transmitter's fingerprint to look like a rogue or unrecognized emitter. The ADC sampling rate and low pass filter (LPF) bandwidth are equally crucial in keeping fingerprint features in the PSD side lobes. Using MicaZ sensors, higher sampling rates retained fingerprint features but increased noise \cite{147}.
 
The transmitter and receiver antenna polarization and orientation can also affect the radiation pattern and fingerprint extraction performance. Emitter antenna hardware imperfections can contribute to the fingerprint feature set for wireless emitter identification \cite{61}. The quantity, type, direction, and polarization of receiving antennas can affect fingerprinting system classification performance.

In supervised learning, several receiver hardware captures for an emitter might be used. The model could generalize and distinguish the emitter fingerprint from recorded waveforms with a bigger training data set. Training on samples from certain receiver hardware and analyzing the learned emitter function can determine fingerprinting algorithm independence.

\subsection{RFFS vulnerabilities} 

Broadcast wireless emitters are vulnerable to identity spoofing. Impersonation, DoS, bandwidth theft, etc. It's often forgotten that passive receiver threats can accumulate cognitive RFFS from transmitter emissions. Another intriguing research challenge to improve wireless security is developing or perturbing the emitter fingerprint so that only legitimate receivers can extract or identify it. RF fingerprinting is expected to be robust to impersonation attacks due to the difficulty of duplicating frontend impairments with replay assaults due to the replaying device's hardware faults. This area is unexplored. Modulation-based RF fingerprinting is more vulnerable to impersonation attempts than transient-based \cite{148}. Another study in \cite{149} examined how numerous low-end receivers made with cheap analog components affected modulation-based RF fingerprinting's impersonation resistance. The receivers' RF fingerprints differed from the transmitter's. As mentioned above, the receiver's hardware defects enhance the fingerprint feature set. They use this information to combat impersonation attacks by claiming that the impersonator would not be able to extract the receiver hardware's fingerprint features, making RFFS even more secure. Jamming DoS assaults, when the intruder transmits continuously on the working frequency, can also affect the RFFS.
RFFS's resilience to DoS attacks will need further investigation. Jamming can also be utilized for clandestine and confidential activities to hide transmitters' RF fingerprints. In \cite{150}, WiFi transmissions were experimentally analyzed for RF fingerprint obfuscation that only the authentic receiver can extract. Random phase faults allow only authentic receivers with a preshared key and randomization index to decode the message and fingerprint.

\section{Conclusion} \label{sec:con}

Deep Learning (DL) has been shown to strengthen RF fingerprinting in identifying a transmitter however, identification accuracy may fail in the presence of unobserved data due to changing channel conditions. Since collecting data under all circumstances is not feasible, data augmentation emerges as a viable approach. Previously, Day-1 data was augmented by using TDL/CDL transformation along with AWGN added on the transformed data. In a four-base station scenario with POWDER dataset, TDL/CDL-based augmentation can result in an accuracy of 77.81\% on Day-2. In this study, we introduce the 5G-only-CDL and WiFi-only-TDL approaches along with the decoupled CDL+TDL augmentation where 5G-waveform data is augmented with CDL transformation, and WiFi-waveform data is augmented with TDL transformation. CDL+TDL,
5G-only-CDL, WiFi-only-TDL augmentation approaches have been shown to increase the accuracy level to 87.59\%, 81.63\%,
79.21\% on Day-2 leading to a remarkable improvement in the performance of the TDL/CDL-based augmentation.



\vspace{12pt}


\begin{IEEEbiographynophoto}
{\bf Omer Melih Gul} (S'17, M'21) received BSc., MSc., and PhD. degrees from the Department of Electrical and Electronics Engineering at Middle East Technical University (METU), Ankara, Turkiye, in 2012, 2014, and 2020, respectively by working as a research assistant at the same department. 
His research interests include machine learning applications, wireless security, networking, scheduling, IoT, UAV, robotics, blockchain and edge/fog computing. 
He has co-authored over 30 papers and 4 book chapters. He was awarded third place in the 2019 Lance Stafford Larson Outstanding Student Paper Award by the IEEE Computer Society.  
He was also awarded third place in the poster competition at the 2021 IEEE Rising Stars Global Conference.
In 2022, he worked as a postdoctoral fellow at the School of Electrical Engineering and Computer Science at the University of Ottawa, Canada. He is a recipient of the best paper award at the 48th Wireless World Research Forum (WWRF) in 2022. 
Currently, he is an assistant professor in the Department of Computer Engineering at Bahcesehir University in Istanbul, Turkiye. He is also an Editor at (Springer) Telecommunications Systems and Wireless Networks journals.

On the IEEE volunteering side, after holding several IEEE volunteer roles, he has currently been a Member-at-Large (MGA Board) and the Region 8 (EMEA) Coordinator at the IEEE Computer Society since April 2021. He was awarded the 2022 IEEE MGA Young Professionals Achievement Award. He was also the chair of IEEE Turkey Young Professionals Affinity Group which was awarded the 2021 IEEE Region 8 Outstanding Young Professionals Affinity Group Award and the 2022 IEEE MGA Young Professionals Hall of Fame Honorable Mention.
\end{IEEEbiographynophoto}

\begin{IEEEbiographynophoto}{\bf Michel Kulhandjian} (M'18-SM'20) received his B.S. degree in Electronics Engineering and Computer Science (Minor), with ``Summa Cum Laude'' from the American University in Cairo (AUC) in 2005, and the M.S. and Ph.D. degrees in Electrical Engineer from the State University of New York at Buffalo in 2007 and 2012, respectively. He was employed at Alcatel-Lucent, in Ottawa, Ontario, in 2012. In the same year he was appointed as a Research Associate at EION Inc. In 2016 has been appointed as a Research Scientist at the School of Electrical Engineering and Computer Science at the University of Ottawa. He was also employed as a senior embedded software engineer at L3Harris Technologies from 2016 to 2021. Currently, he is a Research Scientist at the Electrical and Computer Engineering Department at Rice University. He received the Natural Science and Engineering Research Council of Canada (NSERC) Industrial R\&D Fellowship (IRDF). 
	
His research interests include wireless multiple access communications, adaptive coded modulation, waveform design for overloaded code-division multiplexing applications, RF and audio fingerprinting, channel coding, space-time coding, adaptive multiuser detection, statistical signal processing, machine learning, covert communications, spread-spectrum steganography and steganalysis. He has served as a guest editor for the Journal of Sensor and Actuator Networks (JSAN). He actively serves as a member of Technical Program Committee (TPC) of IEEE WCNC, IEEE GLOBECOM, IEEE ICC, and IEEE VTC. He has served as a Guest Editor for Journal of Sensor and Actuator Networks.
He is a recipient of the best paper award at the 48th Wireless World Research Forum (WWRF) in 2022. 
\end{IEEEbiographynophoto}

\begin{IEEEbiographynophoto}
{\bf Burak Kantarci} (S'05, M'09, SM'12) received the Ph.D degree in computer engineering in 2009. He is a Full Professor and the Founding Director of Smart Connected Vehicles Innovation Centre (SCVIC), and the Founding Director of the Next Generation Communications and Computing Networks (NEXTCON) Research Lab, University of Ottawa. He has coauthored over 250 publications in established journals and conferences, and 15 book chapters. 
He is well known for his contributions to the quantification of data trustworthiness in mobile crowd-sensing (MCS) systems, and game theoretic incentives to promote user participation in MCS campaigns with high value data; as well as AI-backed access control, authentication and machine learning-backed intrusion detection solutions in sensing environments.
In 2022, he has been awarded a Minister's Award of Excellence in Innovation and Entrepreneurship from Ontario Ministry of Colleges and Universities. 
He served as the Chair of IEEE Communications Systems Integration and Modeling Technical Committee, and has served as the Technical Program Co-Chair/Symposium Co-chair of more than twenty international conferences/symposia/workshops, including IEEE Global Communications Conference (GLOBECOM)—Communications Systems QoS, Reliability and Modeling (CQRM) symposium. From 2021-2023, he served as the Secretary of IEEE Social Networks Technical Committee where he is currently the vice-chair. He is an Editor of the IEEE Communications Surveys \& Tutorials, IEEE Internet of Things Journal, Vehicular Communications (Elsevier), and an Associate Editor for IEEE Networking Letters, and Journal of Cybersecurity and Privacy.
He is Editor for several IEEE and Elsevier journals. 
He was an ACM Distinguished Speaker in 2019-2021, currently an ACM Senior Member. 
He is IEEE Systems Council and IEEE Communications Society Distinguished Lecturer.
\end{IEEEbiographynophoto}

\begin{IEEEbiographynophoto}
{\bf Azzedine Touazi}  
received BSc., MSc., and PhD. degrees from Ecole Nationale Polytechnique (ENP) d’Alger and USTHB university, Algeria, all in electrical engineering, in 2003, 2007, and 2017, respectively. His research interests include signal and image processing, machine learning, statistical data analysis, anomaly detection, and 4G/5G  networks. He is currently ML and 4G/5G engineer at ThinkRF Corp. Prior to this he held engineering positions at Alcatel-Lucent and Aerosystems International (ASI). Also, he was a researcher at the Center for Development of Advanced Technologies (CDTA). He is a recipient of the best paper award at 48th Wireless World Research Forum (WWRF) in 2022.
\end{IEEEbiographynophoto}

\begin{IEEEbiographynophoto}
{\bf Cliff Ellement,} Head AI Solutions at ThinkRF, has 25 years of Product and Business development experience. He holds a Masters of Electrical Engineering at Concordia University researching Spread-Spectrum and Mobile communications relating anti-jamming and network optimization techniques. Developed a number of products in the Optical, Microwave, and Data networking businesses at Nortel. Later joining Mitel Networks leading the Unified Communications group developing secure Intelligent Enterprise collaboration applications. Cliff has been involved at the intersection of RF Communications, Security and Artificial Intelligence initiatives to help drive the next generation of wireless security solutions for Critical Infrastructures.
He is a recipient of the best paper award at 48th Wireless World Research Forum (WWRF) in 2022.
\end{IEEEbiographynophoto}

\begin{IEEEbiographynophoto}
{\bf Claude D'Amours} (Member, IEEE)  received the B.A.Sc., M.A.Sc., and Ph.D. degrees in electrical engineering from the University of Ottawa,in 1990, 1992, and 1995, respectively. In 1992,he was employed as a Systems Engineer at Calian Communications Ltd. In 1995, he joined the Communications Research Centre, Ottawa, ON, Canada, as a Systems Engineer. Later in1995, he joined the Department of Electrical and Computer Engineering, Royal Military College of Canada, Kingston, ON, Canada, as an Assistant Professor. He joined the School of Information Technology and Engineering (SITE), which has since been renamed as the School of Electrical Engineering and Computer Science (EECS), University of Ottawa, as an Assistant Professor, in 1999. From 2007 to 2011, he has worked as the Vice Dean of undergraduate studies with the Faculty of Engineering and has been working as the Director of the School of EECS, University of Ottawa, since 2013. His research interests include physical layer technologies for wireless communications systems,notably in multiple access techniques and interference cancellation.
He is a recipient of the best paper award at 48th Wireless World Research Forum (WWRF) in 2022. 
\end{IEEEbiographynophoto}
\end{document}